\documentclass[pre,twocolumn,showpacs,amsmath,amssymb]{revtex4} 

\usepackage{graphicx}       
\usepackage{txfonts}        
\usepackage{mathrsfs}       

\newcommand{\gvec}[1]{\boldsymbol{\mathrm{#1}}}			

\begin{document}


\title{Phase vortices from a Young's three-pinhole interferometer}

\author{Gary Ruben}
\email{gary.ruben@sci.monash.edu.au}
\author{David M. Paganin}
\affiliation{School of Physics, Monash University, Victoria 3800, Australia}

\date{December 18, 2006}

\begin{abstract}
An analysis is presented of the phase vortices generated in the
far field, by an arbitrary arrangement of three monochromatic
point sources of complex spherical waves.  In contrast with the
case of three interfering plane waves, in which an
infinitely-extended vortex lattice is generated, the spherical
sources generate a finite number of phase vortices. Analytical
expressions for the vortex core locations are developed and shown
to have a convenient representation in a discrete parameter space.
Our analysis may be mapped onto the case of a
coherently-illuminated Young's interferometer, in which the
screen is punctured by three rather than two pinholes.
\end{abstract}

\pacs{42.25.-p, 03.65.Vf, 07.60.Ly, 42.25.Hz, 42.50.Dv, 87.80.Cc}      


\maketitle

\section{Introduction}

In a seminal paper, \Citet{Dir31} considered vortical screw-type
dislocations in the phase of complex wavefields, noting the
one-dimensional nature of the associated vortex cores (nodal
lines) in three dimensions (3D). Such phase vortices exist in a
variety of linear and non-linear physical systems that may be
described via complex fields, including the angular-momentum
eigenstates of the hydrogen atom \citep{Mes61a},
the Meissner state of type-II superconductors \citep{Abr57,TilTi90}, vortex states of superfluids \citep{fey55,Pis99} and Bose--Einstein condensates \citep{PitSt03}, optical vortex
solitons \citep{DesTo05}, propagating electron wavefunctions diffracting through crystalline slabs
\citep{AllFa01}, Gaussian random wavefields \citep{Ber78} and optical speckle fields \citep{FreSh93}.

In continuous complex scalar fields, to which the considerations
of the present paper are restricted, vortical behavior is manifest
as screw-type dislocations in the field's multi-valued surfaces of
constant phase \citep{Rie70, NyeBe74, HirGo74}. More precisely, consider a stationary-state,
complex spatial wavefunction or order-parameter field
$\Psi(\gvec{r}) = A(\gvec{r}) \exp \left[i\chi(\gvec{r})\right]$. Here,
$A(\gvec{r})$ is the non-negative real amplitude, $\chi(\gvec{r})$ is
the phase, and $\gvec{r}=(x,y,z)$ is a position vector in 3D.
Note that harmonic time dependence on angular frequency $\omega$ and
time $t$, via the usual multiplicative factor of
$\exp(-i\omega t)$, is suppressed throughout. To determine whether
a phase vortex exists at a point $p$ in a plane $\Pi$ over which
$\Psi(\gvec{r})$ is defined, a line integral of the phase gradient
is evaluated along a smooth infinitesimally-small path $\Gamma$
that encircles $p$. This path is assumed to have a winding number
of unity with respect to $p$, and to be such that the modulus of
$\Psi(\gvec{r})$ is strictly positive at each point on $\Gamma$. One
may then write the following expression for the ``circulation'' of
the field over $\Gamma$ (see, e.g., \Citet{Nye99}):
\begin{equation}
    \oint_\Gamma d\chi = \oint_\Gamma \nabla \chi \cdot \gvec{t} \ ds = 2\pi n.
\end{equation}
Here, $\gvec{t}$ is a unit tangent vector to $\Gamma$, $ds$ is an
infinitesimal line element along $\Gamma$, and $n$ is an integer
corresponding to the number of phase windings about $p$.  Any
non-zero $n$ indicates the presence of a vortex core threading the
path $\Gamma$, with the non-zero value for $n$ being referred to
as its topological charge. The sign of this charge distinguishes
between a vortex ($+$) and an anti-vortex ($-$).

In an optical setting, a common way to generate such phase
vortices is to pass coherent laser or soft X-rays through a spiral
phase plate or forked transmission diffraction grating
\citep{HecMc92, HecMc92a, HeFr95, PeeNu03, OkaSa05, KotKh06}.
\Citet{KimJe97} generated optical vortices with a curved glass plate as
an alternative to the spiral phase plate.
Another means for generating fields with specified vortical phase is a
dynamically computer-controlled, holographic system
\citep{KisYo06}. Other means for creating optical phase
vortices in coherent light include the use of spatial light
modulators \citep{CurGr03}, the use of aberrated lenses to create
vortices in a distorted focal volume \citep{BoiDo67}, and diffraction
from random phase screens \citep{Ber78}.

As an alternative approach, one can forego the use of diffractive
or refractive optical elements, seeking instead to create phase
vortices by the superposition of a small number of non-vortical
fields. For example, \Citet{NicNy87} showed that one can generate
a lattice of vortices by interfering three plane waves. Later,
\Citet{MasDu01} showed that this is a minimum requirement and
reformulated the analysis in terms of phasors.

Here we generalize this idea, by considering the formation of
phase vortices via the superposition of three outgoing spherical
waves, generated by three distinct monochromatic equal-energy
point sources. We see that the resulting system of nodal lines
(vortex cores) exhibits a rich geometry, by developing approximate
analytical expressions for the far-field behaviour of this
nodal-line network.  The three-dimensional space, into which the
sources radiate, is foliated using a family of observation planes
that are parallel to the plane containing the three point sources.
When one observes the wavefield over any such foliating plane, a
2D pattern of point vortices may be seen, the cores of which
coincide with the points at which a nodal line punctures the
plane.

Interestingly, the problem of three interfering spherical waves
may be mapped onto a Young-type experiment, in which a black
screen with three small pinholes is coherently illuminated by a
propagating complex scalar field. Note that this identification is
only possible when one is both sufficiently far from the screen
and sufficiently close to the optic axis, in which case the
radiation transmitted by each of the pinholes is approximately
spherical.

We close this introduction with a brief outline of the remainder
of the paper: We begin by reviewing the manner in which the
superposition of three plane waves may be used to generate an
infinite lattice of phase vortices.  The generalization of this
idea, to the superposition of three outgoing spherical waves, is
then given. We describe the application of a phasor approach to
the spherical-wave arrangement, applying this in the far-field
region asymptotically far from the sources.  Approximate
analytical expressions are derived for the vortex locations. A
representation in terms of a certain parameter space arises,
allowing estimates of the number of vortices and description of a
natural coordinate system for the vortices at the intersections of
a certain family of hyperbolas. A specific case of collinear
sources is explored in detail. We then show how the theory, which
has been derived for spherical point sources, may be mapped onto
the case of a Young's interferometer in which the illuminated
screen contains three rather than two pinholes.


\section{\label{sec:InterferenceOf3PlaneWaves}Phase vortices from the interference of three plane waves}

Consider the following superposition of three planar spatial
wavefunctions:
\begin{equation}
    \label{eqn:planesum}
    \Psi (\gvec{r})= \sum_{j=1}^3 A_j \exp \left[ i\left(\gvec{k}_j \cdot \gvec{r} + \phi_j \right) \right],
\end{equation}
where the non-negative real constants $A_j$ denote the amplitude
of the $j$th wave, $\gvec{k}_j$ are wavevectors corresponding to the
same de Broglie wavelength $\lambda_0=2\pi/|\gvec{k}_j|$, and $\phi_j$
are global phase factors. Notwithstanding the fact that the
constituent plane waves do not have a vortical character, the
above superposition may yield a regular lattice of phase vortices
and anti-vortices \citep{NicNy87, MasDu01}.

\begin{figure}[tb]
    \centering
    \includegraphics[width=85mm]{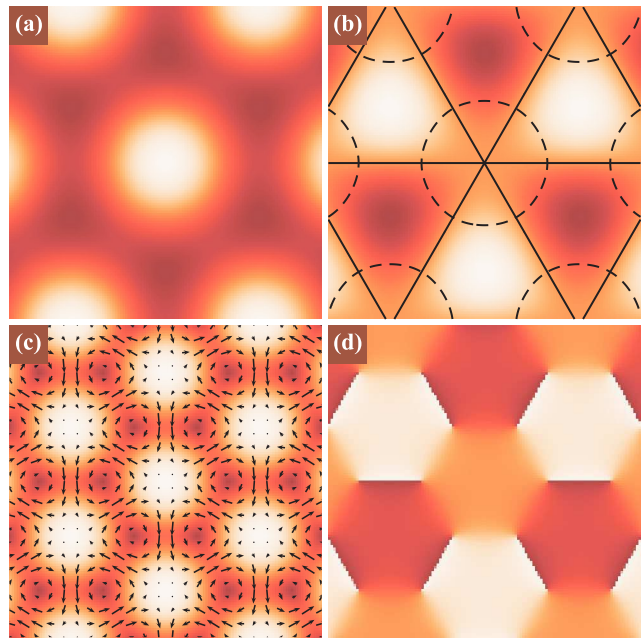}
  \caption{(Color online) Interference of three plane waves giving rise
  to an infinite, regular vortex lattice [see Eq.~\eqref{eqn:planesum}].
  Here the wavevectors are oriented symmetrically with respect to the
  $z$ axis, with $\gvec{k}_1=(\sqrt{3},-1,10), \gvec{k}_2=(-\sqrt{3},-1,10)$ and
  $\gvec{k}_3=(0,2,10)$. The real~(a) and imaginary~(b) parts, amplitude~(c),
  and phase~(d) of the wavefunction $\Psi(x,y,z=0)$ are shown. In~(b), zero
  contours of the real (dashed lines) and imaginary (solid lines) parts
  are overlaid. In~(c) is overlaid the vector field of the probability
  current density, which is seen to rotate counter-clockwise around
  vortices and clockwise around anti-vortices. In~(d), vortices are
  visible as the ends of branch cuts indicated by dark to light
  ($-\pi\rightarrow\pi$) steps. Values in (a-c) are represented by
  linear levels from dark to light (minimum to maximum).}
  \label{fig:planewaves}
\end{figure}

A numerical example of this phenomenon is given in
Fig.~\ref{fig:planewaves}, corresponding to the parameters
$A_1=A_2=A_3=1$ and $\phi_1=\phi_2=\phi_3=0$, with all fields
being evaluated over the plane $z=0$. This example illustrates the
three interfering plane waves giving rise to an infinitely
extended lattice of straight, parallel nodal lines (vortex cores).
These nodal lines intersect the plane $z=0$, to give the location
of the point vortices that are visible as screw dislocations in
the phase map of Fig.~\ref{fig:planewaves}(d). The locations of
these point vortices coincide with both (i) the intersections of
the zero contours in Fig.~\ref{fig:planewaves}(b), and (ii) the
points at which $|\Psi|=0$ in the amplitude plot of
Fig.~\ref{fig:planewaves}(c). Indeed, continuity of the
wavefunction implies that the probability density vanishes at each
vortex core, since these are branch points at which the phase
ceases to be differentiable. The topological charge $n$ of each of
these point defects is seen to be equal to $\pm 1$, as the phase
increases by $\pm 2 \pi$ as one traverses a circuit that encloses
a given vortex core. In this context, we note that higher-charge
vortices are unstable with respect to perturbation \citep{Fre99},
which is why they are not observed in the present setting.

\section{\label{sec:The3SphericalWaveAssembly}Phase vortices from the interference of three spherical waves}


Given that the superposition of three complex plane-wave spatial
wavefunctions may lead to phase vortices \citep{NicNy87, MasDu01}, it is natural to enquire
whether the superposition of three outgoing spherical waves may
not also lead to phase vortices. This latter case is investigated
here.

\subsection{\label{sec:ExtendingThePlaneCase}Extending the plane-wave case to spherical waves}

The complex spatial wavefunction $\Psi_j(\gvec{r})$, due to a point
source at position $\gvec{r}_j$ that is radiating outgoing spherical
waves {\em in vacuo}, is given by
\begin{equation}
    \begin{split}
    \Psi_j(\gvec{r}) &= \frac{A_j}{|\gvec{r}-\gvec{r}_j|} \exp \left[ i \left(k |\gvec{r}-\gvec{r}_j| + \phi_j  \right) \right] \\
                   &\equiv \frac{A_j}{|\gvec{r}-\gvec{r}_j|} \exp \left(i \chi\right),
    \end{split}
\end{equation}
where $k\equiv 2\pi/\lambda_0$.
For all $\gvec{r}\ne \gvec{r}_j$, such spherical waves obey a variety
of linear partial differential equations, including: (i) the
time-independent free-space Schr\"{o}dinger equation for
non-relativistic spinless particles, (ii) the time-independent
free-space Klein-Gordon equation for relativistic spinless
particles, and (iii) the free-space Helmholtz equation for
monochromatic complex scalar electromagnetic waves.  As such, the
following discussions are applicable to all of these physical
systems.

Now consider an assembly of three point sources, all of which have
the same wave-number $k$. Without loss of generality we may
consider these sources to occupy the same plane $z=0$, with source
locations $\gvec{r}_j \equiv (x_j,y_j,0)$, where $j=1,2,3$. The
resulting spatial wavefunction $\Psi(\gvec{r})$ may thus be written
as
\begin{equation}
  \label{eqn:general_sum}
    \Psi(\gvec{r})= \sum_{j=1}^3 \frac{A_j}{|\gvec{r}-\gvec{r}_j|} \exp \left[ i\left(k_j |\gvec{r}-\gvec{r}_j| + \phi_j \right) \right],
\end{equation}
where $|\gvec{r}-\gvec{r}_j| \equiv \sqrt{(x-x_j)^2 + (y-y_j)^2 +
z^2}$ is the distance from the $j{\rm th}$ source to a given
observation point $\gvec{r}$.

\begin{figure}[tb]
    \centering
      \includegraphics[width=55mm]{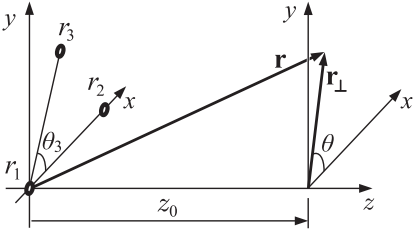}
    \caption{Coordinate system. The point sources $r_1, r_2$, and $r_3$ lie in the plane
    $z=0$. The position vector $\gvec{r}$ is to a point with cylindrical-polar coordinates
    $(r_\perp, \theta, z_0)$.}
    \label{fig:polar_coordinate_system}
\end{figure}

Referring to Fig.~\ref{fig:polar_coordinate_system}, we label both
the $j$th source and its distance from the coordinate origin by
the same symbol $r_j$.
The position vector $\gvec{r}$ and its perpendicular component
$\gvec{r}_\perp$ have lengths $r\equiv|\gvec{r}|$ and
$r_\perp\equiv|\gvec{r}_\perp|$, respectively.

\begin{figure}[tb]
    \centering
    \includegraphics[width=85mm]{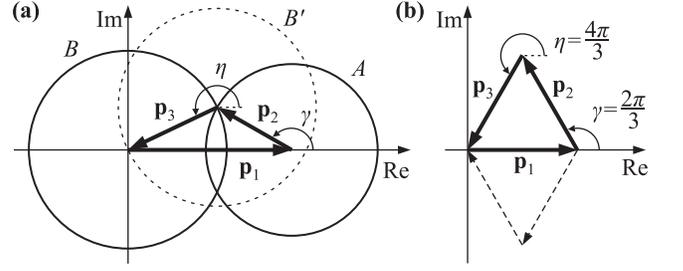}
    \caption{Phasor diagram from which vortex solution conditions are established \citep{MasDu01}.
    Three phasors $\gvec{p}_1, \gvec{p}_2, \gvec{p}_3$ --- which respectively correspond to the
    three terms on the right side of Eq.~\eqref{eqn:general_sum}, at a given point $\gvec{r}$
    on a nodal line --- form a closed triangle corresponding to a resultant zero amplitude. Two
    angles $\gamma$ and $\eta$ arise in the construction, as indicated. (a) In the case
    of arbitrary amplitudes for $\gvec{p}_1$, $\gvec{p}_2$ and $\gvec{p}_3$, the tip of
    $\gvec{p}_2$ is constrained to lie on circle $A$ of radius $|\gvec{p}_2|$, with the tail of
    $\gvec{p}_3$ being constrained to lie on circle $B$ of radius $|\gvec{p}_3|$. (b) For equal
    amplitudes $|\gvec{p}_1|=|\gvec{p}_2|=|\gvec{p}_3|$ an equilateral triangle is formed. The
    dashed construction represents another of the six equivalent alternatives formed by
    permutating the phasor order.}
    \label{fig:phasor}
\end{figure}

To determine the location of the vortices which result from the
superposition of the three spherical waves, we utilize the fact
that vortex cores lie at points of zero amplitude. Given that the
problem is restricted to two degrees of freedom, due to the
complex wavefield representation, a geometric phasor diagram can
be constructed with one phasor for each wave component in
Eq.~\eqref{eqn:general_sum}: see Fig.~\ref{fig:phasor}. We follow
the phasor approach of \Citet{MasDu01} (see also \citep{Pag06}).
At any point coinciding with a vortex core, the phasor components
must sum to zero when placed tip to tail. Note that if the source
amplitudes differ sufficiently, it is possible that no closed
triangle of phasors may be formed, i.e., circles $A$ and $B$ in
Fig.~\ref{fig:phasor} cannot intersect if $|\gvec{p}_1|>|\gvec{p}_2|+|\gvec{p}_3|$. In
this case, no vortices will be produced.

Consider a given point $\gvec{r}$ in space, corresponding to the
special case of Eq.~\eqref{eqn:general_sum} where $A_1=A_2=A_3=1$.
Let $\gvec{p}_1, \gvec{p}_2, \gvec{p}_3$ denote the three complex terms that are summed
on the right side of this equation. These three numbers are
represented as phasors in Fig.~\ref{fig:phasor}. Here, we set
$\phi_1=0$, which implies no loss of generality, since the
invariance of the equations of motion under a shift in the origin
of time implies global phase factors to have no physical meaning.
The circle $B\,'$ represents the possible orientations of $\gvec{p}_3$,
constrained by the tip of $\gvec{p}_2$. The zero sum condition is easier
to construct if $\gvec{p}_3$ is flipped and the circle $B$ drawn
constrained by the tail of $\gvec{p}_1$. The resulting vortex solutions
correspond to the phasors $\gvec{p}_2$ and $\gvec{p}_3$ meeting at the
intersections of the circles $A$ and $B$. Note that there are two
such ``closed triangle'' intersections and hence two apparent
solutions. In the case of equal amplitudes, where
$|\gvec{p}_1|=|\gvec{p}_2|=|\gvec{p}_3|$, symmetry dictates that there are only two
unique solution angles. We consider the triangle in the first
quadrant of the complex plane, which is an arbitrary choice, as
any of the six equivalent constructions formed by permutating the
phasor order will lead to the same solution. Relaxing the
equal-amplitude condition will require consideration of extra
solutions.


\subsection{\label{sec:VorticesInTheFraunhoferRegime}Vortices in the far-field regime}
We evaluate Eq.~\eqref{eqn:general_sum} in the ``far field''
regime, namely in the half-space $z>z_0$ in which $z_0$ is
sufficiently large that
\begin{equation}
        |\gvec{r}-\gvec{r}_j| \sim r - \frac{\gvec{r}_\perp\cdot\gvec{r}_j}{r},
         r\equiv |\gvec{r}| \gg r_j.
\end{equation}
This approximation is applied to the phase term in the exponent of
Eq.~\eqref{eqn:general_sum}. The wavefunction $\Psi$ depends
linearly on the amplitude term $A_j/|\gvec{r}-\gvec{r}_j|$, so for
large $r$ the divisor varies much more slowly with
$|\gvec{r}-\gvec{r}_j|$ than the phase argument. Consequently, the
stronger approximation $|\gvec{r}-\gvec{r}_j| \sim r$ is made to this
term (see, e.g., \citep{Mes61}). Thus, Eq.~\eqref{eqn:general_sum}
becomes
\begin{equation}
  \begin{split}
  \label{eqn:approx_phase}
    \Psi(\gvec{r}) &= \sum_{j=1}^3 \frac{1}{r} \exp \left\{ i \left[k_j \left(r - \frac{\gvec{r}_\perp \cdot\gvec{r}_j}{r}\right) + \phi_j \right] \right\}\\
       &= \frac{1}{r} \exp \left(i k r \right) + \frac{1}{r} \exp \left\{ i \left[k \left(r - \frac{r_\perp r_2}{r} \cos \theta \right) + \phi_2 \right] \right\}\\
       &\ \ \ \ \ \ + \frac{1}{r} \exp \left\{ i \left[k \left(r - \frac{r_\perp r_3}{r} \cos \left(\theta - \theta_3 \right) \right) + \phi_3 \right] \right\},
  \end{split}
\end{equation}
where we have made use of the assumptions that $A_1=A_2=A_3=1$ and that the sources share a single wavenumber $k$.
This expression vanishes when
\begin{equation}
  \begin{split}
    \label{eqn:spherical_phasor}
    1 &+ \exp\left[i\left( {-k\dfrac{r_\perp r_2}{r} \cos{\theta}} + \phi_2 \right)\right] \\
          &+ \exp\left[i\left({-k\dfrac{r_\perp r_3}{r} \cos\left(\theta - \theta_3 \right)} + \phi_3 \right)\right]=0.
  \end{split}
\end{equation}

Geometrically, the above condition reduces to the addition of
three unit-length phasors in the complex plane, such that they
form an equilateral triangle when placed tip-to-tail
\citep{MasDu01}.  This construction is shown in
Fig.~\ref{fig:phasor}(b), with the arguments of the two
exponentials in Eq.~\eqref{eqn:spherical_phasor} being denoted by
$\gamma$ and $\eta$, respectively. These phase angles are uniquely
defined to within an integer multiple of $2\pi$, so that:
\begin{subequations}
    \begin{equation}
        \label{eqn:angle_gamma}
        \gamma = -k\dfrac{r_\perp r_2}{r} \cos \theta + \phi_2 = \dfrac{2\pi}{3} + 2m\pi
    \end{equation}
and
    \begin{equation}
      \label{eqn:angle_eta}
        \eta = -k\dfrac{r_\perp r_3}{r} \cos\left(\theta - \theta_3 \right) + \phi_3 = \dfrac{4\pi}{3} + 2n\pi,
    \end{equation}
\end{subequations}
where $m$ and $n$ are integers.


\subsection{\label{sec:VortexLocation}Vortex locations}
Here, we show how the construction of the previous sub-section can
be used to determine the polar coordinates $r_{\perp mn}$ and
$\theta_{mn}$, of a given point vortex in the plane
$z=z_0$, that is specified by the integer indices $(m,n)$ (cf.
Fig.~\ref{fig:polar_coordinate_system}).

Dividing Eq.~\eqref{eqn:angle_eta} by \eqref{eqn:angle_gamma} gives
\begin{equation}
  \label{eqn:eta_on_gamma}
    \begin{split}
    \dfrac{r_3 \cos\left(\theta - \theta_3 \right)}{r_2 \cos{\theta}}
             &= \dfrac{\dfrac{4\pi}{3} + 2n\pi - \phi_3}{\dfrac{2\pi}{3} + 2m\pi - \phi_2} \\
             &= \dfrac{2(2+3n)\pi - 3\phi_3}{2(1+3m)\pi - 3\phi_2}.
    \end{split}
\end{equation}
We denote the denominator and numerator, on the right-hand side,
as
\begin{subequations}
	\label{eqn:M_and_N}
    \begin{equation}
			\label{eqn:M}
        M(m) \equiv 2(1+3m)\pi - 3\phi_2,
    \end{equation}
and
    \begin{equation}
			\label{eqn:N}
        N(n) \equiv 2(2+3n)\pi - 3\phi_3
    \end{equation}
\end{subequations}
respectively. Next, making the substitution
\begin{equation}
    \begin{split}
    \dfrac{\cos\left(\theta - \theta_3 \right)}{\cos{\theta}}
             &= \dfrac{\cos \theta \cos \theta_3 + \sin \theta \sin \theta_3}{\cos \theta} \\
             &= \cos \theta_3 + \tan \theta \sin \theta_3 \ ,
    \end{split}
\end{equation}
gives
\begin{equation}
  \label{eqn:perturbation_start}
    \dfrac{r_3}{r_2}\left(\vphantom{\dfrac{r_3}{r_2}}\cos \theta_3 + \tan \theta \sin \theta_3\right) = \dfrac{N(n)}{M(m)}.
\end{equation}
Finally, isolating $\theta$ and labelling it with an $mn$
subscript, to identify it with the $(m,n)$th vortex core, gives
the desired expression for the polar angle to the $(m,n)$th vortex
core,
\begin{equation}
  \label{eqn:theta_mn}
    \theta_{mn} = \arctan\left[\dfrac{1}{\sin \theta_3}\left(\dfrac{r_2}{r_3} \dfrac{N(n)}{M(m)} \, -\cos \theta_3 \right)\right].
\end{equation}

With a view to obtaining the radial coordinate $r_{\perp mn}$ of
the vortex core, take Eq.~\eqref{eqn:angle_gamma} and write the
denominator $r$ in terms of its components $z_0$ and $r_\perp$
(see Fig.~\ref{fig:polar_coordinate_system}).  Hence:
\begin{equation}
     -k\dfrac{r_\perp r_2}{\sqrt{{z_0}^2 + {r_\perp}^2}} \cos{\theta_{mn}} = \dfrac{1}{3} M(m).
\end{equation}
Squaring, and then solving for $r_\perp$, we obtain
\begin{equation}
    \label{eqn:r_perp_squared}
    {r_{\perp mn}}^2 = \dfrac{{z_0}^2}{\left(\dfrac{3kr_2 \cos \theta_{mn}}{M(m)} \right)^2-1},
\end{equation}
where an $mn$ subscript has been added to $r_\perp$.
Applying the identity
\begin{equation}
    \label{eqn:cos_to_tan}
    \cos^2 \theta_{mn} = \dfrac{1}{1+\tan^2 \theta_{mn}},
\end{equation}
and making use of Eq.~\eqref{eqn:theta_mn}, we obtain our final
expression for the radial coordinate $r_{\perp mn}$ of the
$(m,n)$th vortex core:
\begin{equation}
  \label{eqn:r_perp}
    r_{\perp mn}  =  \dfrac{\pm z_0}{\sqrt{
                   \dfrac{\big(3kr_2/M(m)\big)^2}{1+\dfrac{1}{\sin^2 \theta_3}\left(\dfrac{r_2}{r_3} \dfrac{N(n)}{M(m)} \, -\cos \theta_3
                   \right)^2}-1}}.
\end{equation}
The positive and negative solutions correspond to two separate
vortices. Note that these coincide with an extra branch of the
$\arctan$ function in Eq.~\eqref{eqn:theta_mn}.  Note, also, that
Eq.~\eqref{eqn:r_perp} is only valid for integers $(m,n)$ that
yield a real number for $r_{\perp mn}$ (cf. Sec.~\ref{sec:ParameterSpace}).

The polar equations~\eqref{eqn:theta_mn} and \eqref{eqn:r_perp}
specify the vortex core locations for all allowed $m$ and $n$
parameter values, in the far-field regime. Note that $r_{\perp
mn}$ is proportional to $z_0$, as one would expect in the
far-field.  This may be contrasted with the case of three
superposed plane waves, where the nodal lines are mutually
parallel \citep{NicNy87, MasDu01}.


\subsection{\label{sec:ParameterSpace}Parameter Space}
For real solutions, the argument of the square root in
Eq.~\eqref{eqn:r_perp} must be positive, imposing a condition on
the allowable $(m,n)$ values for a given source arrangement. In
what follows we set $\phi_j=0$, corresponding to all three point
sources radiating in phase with one another. The integers $m$ and
$n$ must therefore satisfy the inequality
\begin{equation}
  \label{eqn:spherical_mn_constraint}
  \begin{split}
        \left((1+3m) \sin \theta_3 \vphantom{\frac{1}{1}} \right)^2 +
      \left(\left(2+3n \right)\frac{r_2}{r_3}
            -\left(1+3m \right)\cos \theta_3 \right)^2 \\
      < \left(\frac{3kr_2 \sin \theta_3}{2\pi} \right)^2.
  \end{split}
\end{equation}
We claim that this describes the interior of an ellipse in the
Cartesian $(m,n)$ plane, for all non-collinear arrangements of the
three sources.

To prove the above claim, first note that the boundary curve, of
the open region defined by Eq.~\eqref{eqn:spherical_mn_constraint},
is obtained by replacing the inequality in this expression with an
equality.  The resulting equation is consistent with the form of a
general conic section in the $(m,n)$ plane, namely (see,
e.g., \citep{Gib03}):
\begin{equation}
  \label{eqn:general_conic}
    Q(m, n) = am^2+2hmn+bn^2+2gm+2fn+c=0,
\end{equation}
where $a,h,b,g,f$ and $c$ are real numbers given by
\begin{equation}
    \label{eqn:conic_coefficients}
    \begin{split}
        a &= 9 {r_3}^2\\
        h &= -9 r_2 r_3 \cos \theta_3\\
        b &= 9 {r_2}^2\\
        g &= 3 {r_3}^2 - 6 r_2 r_3 \cos \theta_3\\
        f &= 6 {r_2}^2 - 3 r_2 r_3 \cos \theta_3\\
        c &= {r_3}^2 + 4{r_2}^2 - 4 r_2 r_3 \cos \theta_3 - \left(\frac{3k r_2 r_3 \sin \theta_3}{2\pi} \right)^2.
    \end{split}
\end{equation}
Introduce the invariants \citep{Gib03}:
\begin{subequations}
    \label{eqn:conic_invariants}
    \begin{equation}
  \Delta = \left| \begin{array}{ccc} a & h & g \\ h & b & f \\ g & f & c \end{array} \right|,
    \end{equation}
    \begin{equation}
  \delta = \left| \begin{array}{ccc} a & h \\ h & b \end{array}
  \right|,
    \end{equation}
and
    \begin{equation}
  \tau = a+b.
    \end{equation}
\end{subequations}
Substituting Eqs~\eqref{eqn:conic_coefficients} into
Eqs~\eqref{eqn:conic_invariants} and evaluating gives
\begin{subequations}
    \begin{equation}
        \label{eqn:invariantBigDelta}
  \Delta = -\left(\frac{27 k}{2\pi} \left(r_2 r_3 \sin \theta_3 \right)^2 \right)^2,
    \end{equation}
    \begin{equation}
        \label{eqn:invariantSmallDelta}
  \delta = \left(9 r_2 r_3 \sin \theta_3 \right)^2,
    \end{equation}
and
    \begin{equation}
  \tau = 9 \left( {r_2}^2 + {r_3}^2 \right).
    \end{equation}
\end{subequations}
In order for Eq.~\eqref{eqn:general_conic} to correspond to
an ellipse, the discriminant conditions $\Delta \neq 0$, $\delta > 0$
and $\Delta / \tau < 0$ must be satisfied.
These three conditions are met when: (i) $r_2,r_3 \neq 0$, and
(ii) $\theta_3 \neq p\pi, \ p \in \mathbb Z$.  This will always be
true for non-collinear arrangements of three distinct sources.
Since the area of the corresponding ellipse is finite, for three
non-collinear sources each of which are separated by a finite
distance, we have a finite number of vortices labelled by the
integer pairs $(m,n)$ obeying
Eq.~\eqref{eqn:spherical_mn_constraint}.

\begin{figure}[tb]
    \centering
        \includegraphics[width=62mm]{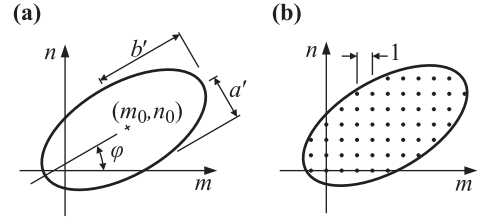}
    \caption[Parameter space ellipse.]{A parameter-space ellipse,
    in the $(m,n)$ plane, sets an upper limit on the number of vortices that are created by the
    interfering radiation from three point sources. (a) The bounding ellipse has
    center $(m_0,n_0)$, makes an angle $\varphi$ to the positive $m$-axis
    and has semi-axis lengths $a'$ and $b'$. (b) The enclosed $(m,n)$
    pairs lie on a discrete square lattice of unit spacing. Each of these interior
    points corresponds to two vortices.}
    \label{fig:ellipse}
\end{figure}

The parameter-space ellipse has center $(m_0,n_0)$, rotated
anti-clockwise at an angle $\varphi$, with semi-axis lengths $a'$
and $b'$ (Fig.~\ref{fig:ellipse}).
These ellipse parameters are given by well known expressions in
terms of the coefficients in Eq.~\eqref{eqn:conic_coefficients}.
The center and rotation angle are given by \citep{Gib03}
\begin{equation}\label{eqn:ellipse_centre_and_rotation_angle}
    \begin{split}
    (m_0, n_0) &= \left( \frac{bg-hf}{h^2-ab}, \frac{af-hg}{h^2-ab} \right)
                          = \left( - \frac{1}{3}, - \frac{2}{3} \right), \\
  \varphi &= \frac{1}{2}{{\rm arccot}\left(\frac{b-a}{2h}\right)} = \frac{1}{2}{{\rm arccot}\left({\frac{{r_3}^2-{r_2}^2}
           {2 r_2 r_3 \cos \theta_3}}\right)}.
    \end{split}
\end{equation}
The semi-axis lengths $a' \equiv s_+$ and $b' \equiv s_-$ are
given by
\begin{equation}
    \label{eqn:semiaxes}
    s_{\pm}=\sqrt{\left| \frac{\Delta}{\lambda_{\pm}\delta}
    \right|}\ ,
\end{equation}
where $\lambda_{\pm}$ denotes the two solutions to the quadratic
\begin{equation}
        \lambda^2 - \tau\lambda + \delta = 0.
\end{equation}
Thus
\begin{equation}
        \lambda_{\pm} = \frac{9}{2} \left( {r_2}^2 + {r_3}^2 \pm
            \sqrt{ \left(2 r_2 r_3 \cos \theta_3 \right)^2 + \left({r_2}^2 - {r_3}^2 \right)^2}
            \right),
\end{equation}
so that the semi-axis lengths are given by
\begin{equation}
    \label{eqn:semiaxis_solutions}
    s_\pm=\frac{|k r_2 r_3 \sin \theta_3|}{\pi \sqrt{2\left| \sqrt{\left(2 r_2 r_3 \cos \theta_3 \right)^2 + \left( {r_2}^2 - {r_3}^2 \right)^2}\pm \left( {r_2}^2 + {r_3}^2 \right)
    \right|}}.
\end{equation}

For fixed $k$, $r_2$, and $r_3$, and variable $\theta_3$, Eqs
\eqref{eqn:ellipse_centre_and_rotation_angle} and
\eqref{eqn:semiaxis_solutions} define a family of ellipses. A
``bounding rectangle'' may be constructed, as the envelope of this
continuum of ellipses. Any one ellipse in this family,
corresponding to a particular value of $\theta_3$, touches each
side of this bounding rectangle exactly once. The bounding
rectangle is centered at $(m_0,n_0)$, and has dimensions of
$kr_2/\pi$ and $kr_3/\pi$ in the $m$ and $n$ directions,
respectively. (Note that these dimensions are found by setting
$\theta_3=\pi/2$ in Eq.~\eqref{eqn:semiaxis_solutions}.) As
$\theta_3$ is varied from 0 to $\pi$, while keeping $k$, $r_2$,
and $r_3$ fixed, the parameter-space ellipse transforms from: (i)
a line at $45^\circ$ to the $m$-axis, identified with the
positive-gradient diagonal to the bounding rectangle, to (ii) a
series of non-degenerate ellipses, each of which touch each side
of the bounding rectangle exactly once, to (iii) a line at
$-45^\circ$, namely the negative-gradient diagonal to the bounding
rectangle. Note that no vortices are produced in the limit cases
(i) and (iii) above, since the open region bounded by a straight
line is an empty set (cf. Eq.~\eqref{eqn:spherical_mn_constraint}).
The change in shape of the ellipse with $\theta_3$ is symmetric
about $\theta_3=\pi$ (corresponding to case (iii)), so that the
ellipse for $\theta_3=\pi-\xi$ is coincident with that for $\pi+\xi$, for any angle $\xi$.

\begin{figure}[tb]
  \centering
  \includegraphics[width=85mm]{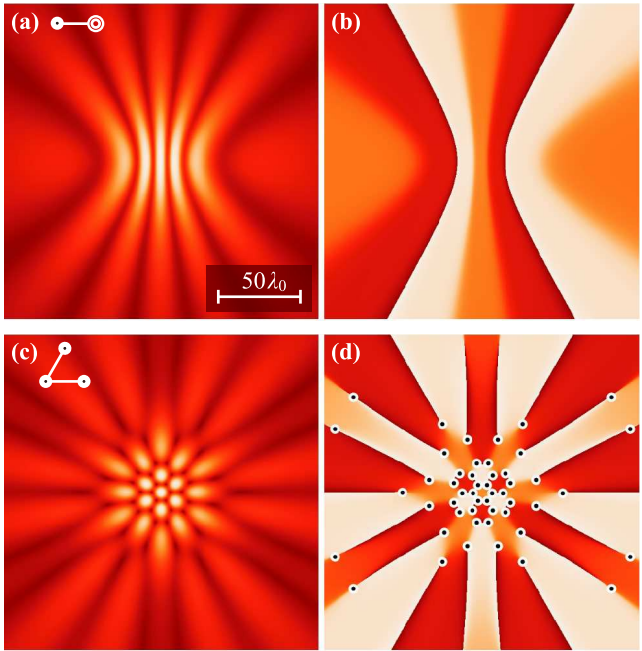}
  \caption{(Color online) (a,c) Amplitude $|\Psi|$ and (b,d) associated
  phase $\chi$ from Eq.~\eqref{eqn:general_sum} at $z_0\!\!=\!\!25\lambda_0$ with different source arrangements, shown pictographically alongside the (a) and (c) labels. Small circles overlaid on numerical simulations show the vortex locations as approximated by Eqs~\eqref{eqn:theta_mn} and \eqref{eqn:r_perp}. Values in (a, c) are represented by linear levels from dark to light (minimum to maximum). The phase ranges from $-\pi$ (dark) to $\pi$ (light). All sources are in phase (i.e. $\phi_2\!\!=\!\!\phi_3\!\!=\!\!0$), $\lambda_0\!\!=\!\!\pi$ and the source arrangements are:
(a,b)~$r_2\!\!=\!\!r_3\!\!=\!\!3\lambda_0,~\theta_3\!\!=\!\!0^\circ$.
(c,d)~$r_2\!\!=\!\!r_3\!\!=\!\!3\lambda_0,~\theta_3\!\!=\!\!60^\circ$. Note
that a spherical background has been subtracted from all phase
maps, as described in the main text.
}
  \label{fig:examples}
\end{figure}

\begin{figure}[tb]
  \centering
    \includegraphics[width=85mm]{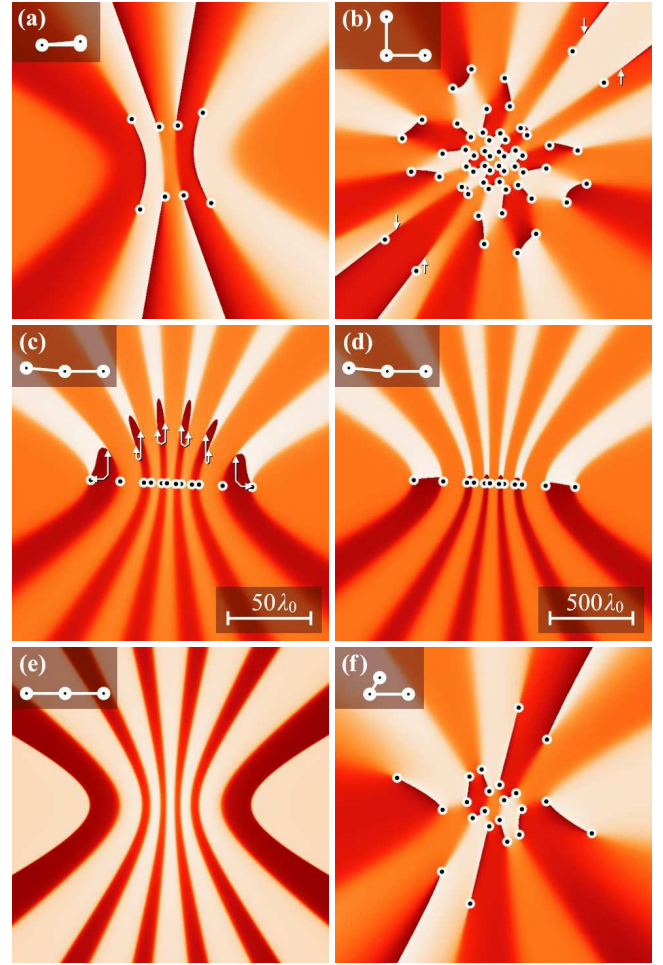}
  \caption{(Color online) Phase $\chi$ from Eq.~\eqref{eqn:general_sum} and vortex-core locations approximated by Eqs~\eqref{eqn:theta_mn} and \eqref{eqn:r_perp}, with variation in source arrangement. The representation and parameters are as described in Fig.~\ref{fig:examples}. The scale-bar in (c) applies to all images and corresponds to $z_0\!\!=\!\!25\lambda_0$, except for (d) which corresponds to $z_0\!\!=\!\!250\lambda_0$. Small arrows in (b,c) point to the exact vortex locations, to indicate their deviation from the far-field predictions.  $\lambda_0\!\!=\!\!\pi$ and source arrangements are:
(a)~$r_2\!\!=\!\!r_3\!\!=\!\!3\lambda_0,~\theta_3\!\!=\!\!10^\circ$.
(b)~$r_2\!\!=\!\!r_3\!\!=\!\!3\lambda_0,~\theta_3\!\!=\!\!90^\circ$.
(c)~$r_2\!\!=\!\!r_3\!\!=\!\!3\lambda_0,~\theta_3\!\!=\!\!175^\circ$.
(d)~$r_2\!\!=\!\!r_3\!\!=\!\!3\lambda_0,~\theta_3\!\!=\!\!175^\circ,~z_0\!\!=\!\!250\lambda_0$.
(e)~$r_2\!\!=\!\!r_3\!\!=\!\!3\lambda_0,~\theta_3\!\!=\!\!180^\circ$.
(f)~$r_2\!\!=\!\!3\lambda_0,r_3\!\!=\!\!1.5\lambda_0,~\theta_3\!\!=\!\!60^\circ$. A
spherical background has been subtracted from all phase maps (see
main text).}
  \label{fig:phases}
\end{figure}

\begin{figure}[tb]
  \centering
    \includegraphics[width=85mm]{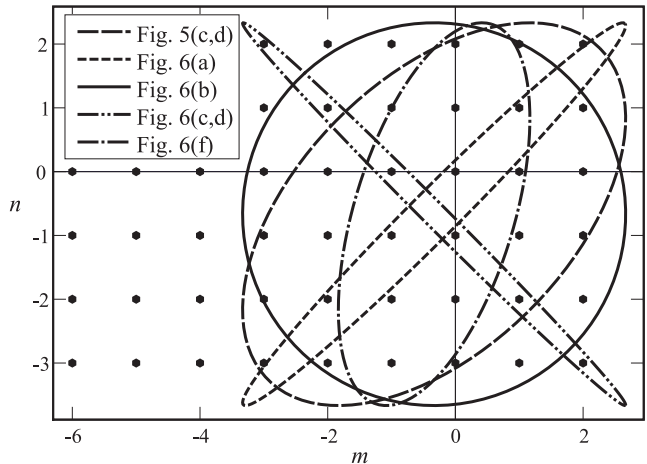}
  \caption{Parameter space ellipses corresponding to Figs~\ref{fig:examples} and \ref{fig:phases}. Enclosed lattice points define vortex locations.}
  \label{fig:parameter_space}
\end{figure}

Figs~\ref{fig:examples} and \ref{fig:phases} present simulations
with various source geometries showing predicted vortex locations
for the far-field case. The corresponding parameter
space ellipses are shown in Fig.~\ref{fig:parameter_space}.
The fields of view of the intensity and phase plots do not show the
outermost vortex cores in most cases. This is evident from a count
of the lattice points enclosed by the corresponding ellipse in
Fig.~\ref{fig:parameter_space}. Only Figs~\ref{fig:phases}(c) and
(d) have all the generated vortices within the visible region. The
corresponding parameter-space ellipse, near $-45^\circ$ in
Fig.~\ref{fig:parameter_space}, encloses 6 lattice points --- this
corresponds to the $6\times 2 = 12$ phase vortices in
Figs~\ref{fig:phases}(c) and (d).

Note that, to aid visualization in all of the phase plots in
Figs~\ref{fig:examples} and \ref{fig:phases}, a constant
spherical background has been subtracted. Indeed, far from the
three point sources, one may meaningfully write the wavefunction
as a single expanding ``background'' spherical wave, multiplied by
an envelope whose functional form depends on the particular local
arrangement of the sources. By subtracting the phase of this
spherical background from all of the displayed phase maps, the
structure of the envelope alone --- including any vortical
structure --- may be examined, without the distraction of a large
number of concentric phase contours from the background wave. Note
also that this background spherical wave corresponds to an
effective source located at the geometric centroid of the real
sources, with amplitude $A=A_1+A_2+A_3$.

Return consideration to Fig.~\ref{fig:examples}(a), which exhibits
a degenerate case in which two of the point sources are
co-located, thereby reducing the system to two in-phase spherical
sources with one having twice the amplitude of the other. As
expected for only two sources, no vortices are generated. Rather,
one has a series of Young-type fringes, with minima corresponding
to a series of nodal planes in 3D. Each of these nodal planes is
seen to constitute a ``domain wall'' for the phase \citep{Pis99}
[see Fig.~\ref{fig:examples}(b)]. The phase is
not left-right mirror symmetric, due to the positioning of the
geometric centroid for three sources being to the right of center.
Thus, there is a global tilt to the phase, which is observed to
cycle through two branch-cuts. For this example, the
parameter-space ellipse (not shown) is the limiting case of a line
at $45^\circ$. The complementary limiting case is
Fig.~\ref{fig:phases}(e) where the three sources are collinear;
the ellipse (not shown) is a line at $-45^\circ$ and again no
vortices are generated. The phase here is mirror symmetric since:
(i) the initial field configuration is mirror symmetric, and (ii) the
wavefield propagator is rotationally symmetric. In
Fig.~\ref{fig:phases}(a) one of the sources has been moved just
enough from coincidence with another so that the corresponding
ellipse just encompasses some lattice points and vortices are
created. There are four points clearly within the ellipse, giving
rise to the vortices seen in the panel. If the additional two points
near the boundary are just inside the ellipse, more
vortices will be present at large $r_\perp$ and these will lie at
the other ends of the branch-cut lines; otherwise the branch-cut
lines will extend to infinity. Fig.~\ref{fig:phases}(b) shows the
case for $\theta_3=90^\circ$, giving a circle in parameter space.
The number of vortices will therefore be close to the maximum for
fixed values of $r_2$ and $r_3$.

The far-field solution is usually considered valid for a Fresnel
number $N_F\equiv a^2/\lambda_0 z \ll 1$, where $a$ is the largest
transverse length scale present in the system (see, e.g., \citep{Pag06}). We
take this to be the maximum pinhole--pinhole spacing. The
far-field condition corresponding to Fig.~\ref{fig:examples} is
then $(3\lambda_0)^2/\lambda_0 z \ll 1$ or, say $z > 100\lambda_0$. The
value of $z_0$ for numerical simulations in
Fig.~\ref{fig:phases} was deliberately chosen as $25\lambda_0$ to
show up visual disagreements between the vortices in the
numerically determined phase map and their analytically determined
locations. Small arrows illustrate the true vortex positions for
some vortices which do not coincide with their predicted far-field
positions. The discrepancy between numerical and analytical
results is highlighted in Fig.~\ref{fig:phases}(c), a case arising
from all parameter space lattice points lying close to the ellipse
boundary. In Fig.~\ref{fig:phases}(d), the Fresnel number is
smaller and the correspondence is improved, but for any finite
$z_0$, there will be some angle, close to $\pi$, at which the
numerical and analytical prediction disagree by an arbitrarily
large amount. Predictions are more reliable for $(m,n)$ lattice
points closer to the ellipse center. In contrast with the
far-field prediction that vortices may abruptly appear and
disappear with infinitesimal changes in source arrangement as
lattice points cross the ellipse boundary, we observe through
simulations at finite $z_0$, that the vortex location becomes
highly sensitive to source arrangement, with vortices arriving
from and escaping transversely to an infinite $r_\perp$ distance.
This point is further explored in Sec.~\ref{sec:ExplorationOfParameterSpace}.


\subsection{\label{sec:SourcePhaseVariation}Source phase variation}
We now consider the effects of source phase variation on the geometric parameter space description.

The functions $M(m)$ and $N(n)$ [Eqs~\eqref{eqn:M_and_N}] contain the vortex coordinates' dependency on source phase via Eqs~\eqref{eqn:theta_mn} and \eqref{eqn:r_perp}, where $\phi_j$ is the relative phase of the $j$th source and $\phi_1=0$. These may be rewritten as
\begin{subequations}
	\begin{equation}
		M(m) \equiv 2\pi \left[1+3\left(m - \phi_2/{2\pi} \right)\right]
	\end{equation}
and
	\begin{equation}
		N(n) \equiv 2\pi \left[2+3\left(n - \phi_3/{2\pi} \right)\right],
	\end{equation}
\end{subequations}
to highlight the $2\pi$ phase periodicity.

We see that a $2\pi$ change may be absorbed as an integer change in an associated parameter-space coordinate $m$ or $n$. The subtraction of a value from $m$ or $n$ corresponds geometrically to a translation of the $(m,n)$ lattice points along the associated axis, with a change of $2\pi$ effecting a translation of one lattice unit.

In summary, parameters defining the source configuration or the wave-number map to the parameter space as different ellipse constructions, whereas source phase variations correspond to translations of the lattice itself.


\subsection{\label{sec:EstimateOfNumberOfVortices}Estimate of number of vortices}
In contrast to the case of three interfering plane waves reviewed
in Sec.~\ref{sec:InterferenceOf3PlaneWaves}, in which
infinitely many nodal lines are produced, the analysis of the
preceding sub-section implies that only a finite number of
vortices are produced by three overlapping spherical waves,
provided that there is a finite spacing between the three
corresponding point sources. Here, we give a simple means to
estimate the number of nodal lines, as a function of the geometry
of the three point sources.

As mentioned earlier, the two-valued nature of
Eq.~\eqref{eqn:r_perp} implies that each $(m,n)$ pair gives rise
to a pair of vortices. Because the $m$ and $n$ values have unit
spacing [see Fig.~\ref{fig:ellipse}(b)], the number of vortices
$n_v$ may be approximated by twice the area $\pi a' b'$ of the
ellipse. Thus
\begin{equation}
    n_v \approx 2\pi a' b',
\end{equation}
or
\begin{equation}
    \label{eqn:vortex_number}
    n_v \approx \frac{k^2}{2\pi} \Big| r_2 r_3 \sin \theta_3 \Big|=\frac{k^2\mathscr{A}}{\pi},
\end{equation}
where $\mathscr{A}$ is the area of a triangle whose vertices
coincide with the locations of the three point sources.

One may ask whether it is possible to develop an exact expression for the number of
lattice points enclosed by our parameter-space ellipse, thereby
improving on the approximation for $n_v$ given in
Eq.~\eqref{eqn:vortex_number}. Indeed, this question is addressed by a famous problem in number theory known as ``Gauss's circle problem''. With $\phi_2=2\pi/3$, $\phi_3=-2\pi/3$ and $\theta_3=\pi/2$, the conditions corresponding to the common form of Gauss's circle problem are established; the ellipse is a circle and is centered on a lattice point. The resulting solution for this case is both involved and well known, and so will not be given here (see, e.g., \Citet{And94}).
Instead, we merely note that this establishes an unexpected and beautiful connection, between the physical system considered here, and a certain problem in the theory of numbers.


\subsection{\label{sec:VorticesObtainedFromHyperbolas}Vortex trajectories from phase variation}
In most of the preceding discussion, the phases
$\phi_1,\phi_2,\phi_3$ of the sources were all set to zero.  Here,
we consider how vortices move in response to varying the phase of
one of the sources.

The equations of these curves are found by eliminating the phase
$\phi_j$ corresponding to the source $r_j$ in
Eqs~\eqref{eqn:theta_mn} and \eqref{eqn:r_perp}. For example, the
trajectories for variation of source $r_2$ are found by solving
Eq.~\eqref{eqn:theta_mn} for $\phi_2$ and substituting this into
Eq.~\eqref{eqn:r_perp}. Repeating this for $\phi_3$ gives a second
set of trajectories along which the vortices move with $\phi_3$
variation. The resulting equations are
\begin{subequations}
    \label{eqn:hyperbolas}
    \begin{equation}
        \label{eqn:m_hyperbolas}
        r_{\perp m} = \frac{M(m) \ z_0}{\sqrt{(3 k r_2 \cos \theta_m)^2 - M(m)^2}}
    \end{equation}
and
    \begin{equation}
        \label{eqn:n_hyperbolas}
        r_{\perp n} = \frac{N(n) \ z_0}{\sqrt{(3 k r_3 \cos (\theta_n-\theta_3))^2 - N(n)^2}},
    \end{equation}
\end{subequations}
where the subscripts $mn$ have been changed to $m$ or $n$ to
highlight the independence of the equations with respect to the
complementary parameter space coordinate. Note that
Eq.~\eqref{eqn:m_hyperbolas} is the same as
Eq.~\eqref{eqn:r_perp_squared}, with minor manipulation.

\begin{figure}[tb]
    \centering
    \includegraphics[width=56mm]{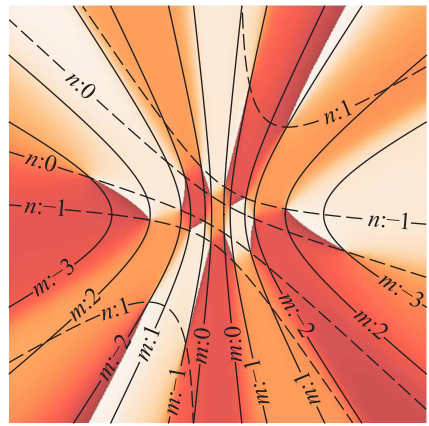}
  \caption{(Color online) Vortices move with varying source phase along trajectories and lie at the intersections of two independent sets of hyperbolas. Two independent phases give rise to two sets of hyperbolas indexed by $m$ (solid lines) and $n$ (dashed lines) respectively. The hyperbolas are shown overlaying phase $\chi$ from Eq.~\eqref{eqn:approx_phase}. Note that the result of this equation is $\ge 0$ for arbitrary $(r_\perp, \theta)$. The parameters are the same as Fig.~\ref{fig:phases}(f).}
  \label{fig:trajectories}
\end{figure}

Fig.~\ref{fig:trajectories} shows the trajectories overlaying the numerical simulation results for the source arrangement seen in Fig.~\ref{fig:phases}(f).

We have seen (Sec.~\ref{sec:SourcePhaseVariation}) that variation of $\phi_2$ (or $\phi_3$) corresponds to a translation of the $m$ (or $n$) lattice alone. Thus, our choice to eliminate the respective phase has led to the separation of the $m$ and $n$ dependence. The equations describe hyperbolas, where the polar coordinate system origin is centred between the two hyperbola branches. Note that this polar form is less common than that typically seen in the study of conics, in which the origin is placed at the focus of one branch. Eqs~\eqref{eqn:hyperbolas} may be thought of as a (one-to-many) mapping from lines of constant $m$ or $n$ in parameter space to hyperbolas in real space.

The angle $\theta_3$ in Eq.~\eqref{eqn:n_hyperbolas} corresponds to the rotation of the $n$-indexed hyperbola axes with respect to the coordinate system axes. The square-root limits the range of real solutions, hence the number of hyperbolas and the number of vortices. The domain of the parameter-space variable $m$ (or $n$) is restricted to match that determined by the parameter-space ellipse. Thus these equations may be applied to determine the valid domains of $m$ or $n$ independently, without recourse to the ellipse solution.

Fixing the source phases defines two particular hyperbolas corresponding to the parameter space coordinates $m$ and $n$ respectively, with vortices located at the intersections of an $m$ and an $n$ curve. Both branches are included, giving four separate curves. Where one branch of an $m$ (or $n$) hyperbola intersects both branches of an $n$ (or $m$) hyperbola, only one intersection is observed to correspond to a physical vortex solution.

These equations also describe an intersecting curvilinear coordinate system. The location of a vortex on the phase map may be thought of as addressable by selection of a particular patch having coordinates $(m,n)$ and then finely addressable within that patch by variation of the source phases. Variation of the phases from 0 to $2\pi$ shifts the vortices along one of the two families of curvilinear trajectories within the patch. When there are a moderate or large number of vortices present, the patches in the central region of the phase map are correspondingly small. In this case, the vortices shift a small distance with changing phase. In the outer, sparsely populated regions, small changes in phase lead to arbitrarily large distance changes since the outermost patches have boundaries at infinity.

It is not unexpected that vortices should lie along intersecting hyperbolas. Two point sources naturally give rise to surfaces of constant phase that are hyperboloids of two sheets having the sources as their foci. The difference in path length being constant establishes the condition for constant phase and is equivalent to the well known geometric construction method for hyperboloids. The intersections of hyperboloids of two sheets with the $z=z_0$ plane will be hyperbolas of two branches. These will not have phase values at the amplitude maxima or minima due to any individual pair, but will be at some other value $\chi=a+ib$ on one set and $-\chi=-a-ib$ on the other, summing to zero at the vortex locations.


\subsection{\label{sec:ExplorationOfParameterSpace}The parameter space near collinearity}

Here we examine the behaviour of the nodal lines as $\theta_3$
approaches $\pi$ (three collinear sources) when $\phi_2=\phi_3=0$,
since we observe a marked deviation between numerical simulation
and analytical results in this regime [see
Fig.~\ref{fig:phases}(c), which clearly illustrates this
discrepancy].

An analysis around $\theta_3=\pi$ may be performed by substituting
$\theta_3=\pi+\varepsilon$ into
Eq.~\eqref{eqn:perturbation_start}, where $\varepsilon$ is small:
\begin{equation}
    \label{eqn:perturbation_relation}
    \dfrac{r_3}{r_2}\left(\vphantom{\dfrac{r_3}{r_2}}\cos (\pi+\varepsilon) + \tan \theta \sin (\pi+\varepsilon)\right) = \dfrac{N(n)}{M(m)}.
\end{equation}
Applying the small-angle approximations $\cos\varepsilon\sim 1$,
$\sin\varepsilon\sim \varepsilon$,
Eq.~\eqref{eqn:perturbation_relation} yields
\begin{equation}
    \label{eqn:perturbation_angle}
    \tan\theta_{mn}(\varepsilon) = \dfrac{1}{\varepsilon}\left(\dfrac{r_2}{r_3} \dfrac{N(n)}{M(m)} + 1
    \right)
\end{equation}
for the $\theta_{mn}$ coordinate of the vortex core. Substituting
into Eq.~\eqref{eqn:r_perp_squared} and applying
Eq.~\eqref{eqn:cos_to_tan} gives $r_{\perp mn}$ for our special
case
\begin{equation}
    {r_{\perp mn}}^2 = \dfrac{{z_0}^2}{\left(\dfrac{3kr_2 }{M(m)} \right)^2 \dfrac{1}{1+\tan^2 \theta_{mn}}-1}.
\end{equation}
Finally, substituting Eq.~\eqref{eqn:perturbation_angle}, we get
\begin{equation}
    r_{\perp mn}(\varepsilon) = \pm\dfrac{z_0}{\sqrt{ \dfrac{\big(3kr_2/M(m)\big)^2}{1+\dfrac{1}{\varepsilon^2}\left(\dfrac{r_2}{r_3} \dfrac{N(n)}{M(m)} + 1 \right)^2}-1}}.
\end{equation}

In the limit $\varepsilon\rightarrow 0$, we now see that
$\theta_{mn}(\varepsilon) \rightarrow \pi/2$ and $r_{\perp
mn}(\varepsilon) \rightarrow \mp i z_0$. Regarding the latter
limit, $r_{\perp mn}(\varepsilon)$ first approaches infinity and
then becomes imaginary.

We may widen the context of this result by realizing that it is but one
example of a lattice point crossing an ellipse boundary. Whenever this
occurs, the far-field prediction is that associated vortices will be created or
destroyed instantaneously. In contrast, in numerical simulations,
which are at some finite distance from the source, vortices
rapidly enter from or escape to infinity.


\subsection{\label{sec:Pinholes}Relation to the Young's three-pinhole interferometer}
Here we show how the to map our results for three spherical point
sources, hitherto the main subject of this paper, onto a
three-pinhole Young's interferometer. In this interferometer,
coherent radiation illuminates a black screen that is punctured
with three small pinholes, with the resulting transmitted
radiation being observed at a distance that is large compared to
the spacing between the pinholes. Note that the assumptions of equal
amplitude and a single wavenumber, applied in Eq.~\eqref{eqn:approx_phase},
correspond to uniform illumination of the screen. The following argument is based
on the Rayleigh--Sommerfeld diffraction theory (see, e.g., \citep{BorWo99}).

The Rayleigh--Sommerfeld diffraction integral of the first kind
yields a rigorous solution to the time-independent Schr\"{o}dinger
equation (Helmholtz equation) in a vacuum-filled half-space, for a
field that obeys the Sommerfeld radiation condition. For a field
$U^{(i)}$ incident on an aperture $A$ centered on the plane $z=0$,
the wavefunction $U$ may be determined at an arbitrary point
$(x,y,z)$ in the half-space $z\ge 0$. With the boundary conditions
that $U(x,y,z=0)\approx U^{(i)}(x,y,z=0)$ when $(x,y,z=0)$ is in
$A$, and $U(x,y,z=0)\approx 0$ when $(x,y,z=0)$ is not in $A$, the
diffraction integral reads:
\begin{equation}
    \label{eqn:rayleigh_sommerfeld_integral}
    U(x,y,z) = \int\!\!\!\int_A U^{(i)}(x',y',0) \ K(x,y,x',y') \ dx' dy',
\end{equation}
where
\begin{equation}
    \label{eqn:propagator}
    K\equiv\frac{1}{2\pi} \frac{\partial}{\partial z} \left(\frac{\exp \left(ikr\right)}{r}\right)
\end{equation}
is a propagator and $r=\sqrt{(x-x')^2+(y-y')^2+z^2}$. Evaluation
of the derivative gives
\begin{equation}
    \label{eqn:rayleigh_sommerfeld_field}
    K = \frac{1}{2\pi}\frac{z}{r}\left(\frac{ik}{r} - \frac{1}{r^2} \right) \exp \left(ikr\right).
\end{equation}

For a pinhole aperture, $U^{(i)} \equiv \delta(x-x',y-y',z)$, and
the wavefield $U(x,y,z)$ takes the form of the propagator,
Eq.~\eqref{eqn:rayleigh_sommerfeld_field}. In general, the pinhole
aperture does not produce spherical waves.  However, when the
observation point is such that $r \gg \lambda_0$, the first term in
parentheses dominates the second, giving
\begin{equation}
    \label{eqn:fraunhoferpropagator}
    K \approx \frac{ikz}{2\pi r^2}\exp \left(ikr\right).
\end{equation}

It can now be seen that Eq.~\eqref{eqn:spherical_phasor}, which
resulted from factoring out a common $\exp \left(ikr\right)/r$
term, would be unchanged by instead factoring out a common term of
$ikz\exp \left(ikr\right)/2\pi r^2$. Similarly, when one is both
in the far-field and close to the $z$-axis, $z\approx r$, so
Eq.~\eqref{eqn:fraunhoferpropagator} reduces to a spherical
wavefunction with a constant multiplier, as asserted.

In summary, the analysis of the preceding sections may be
mapped onto the case of a Young's three-pinhole interferometer,
in the far field. The nodal planes of the two-pinhole Young's
experiment, which are unstable with respect to perturbations,
therefore decay into a nodal-line network of vortex cores when the
third pinhole is added.


\section{\label{sec:Summary}Conclusion}

A network of phase vortices was seen to be generated by the
superposition of three stationary-state sources of outgoing
spherical waves. We presented an analysis of the structure of the
associated vortex cores (nodal lines), in the far-field regime. A
finite number of vortices was seen to be generated. Determination
of the number of such vortices was mapped onto the problem of
determining how many points, on a two-dimensional cubic lattice,
lie within a given ellipse. The equation of the ellipse
depends in a known way on the geometry of the sources. The
parameter space description also gives some insight into the
effects of varying both the arrangement of the three sources, and
their relative phase.  Indeed, phase variation of two of the
sources provides a means for precisely positioning one or several
vortex cores.  Lastly, we showed how to map all of
the preceding analyses onto the problem of determining the
far-field disturbance that results when a three-pinhole Young's
interferometer is coherently illuminated.  In contrast to the
classical two-pinhole Young's interferometer, in which the resulting
diffracted field vanishes over a series of nodal planes, the
three-pinhole interferometer yields a quite different phase
topology, permeated with a rich structure of nodal
lines that thread vortex cores.

\acknowledgments{The authors wish to thank M.J. Morgan for many
fruitful discussions and insights related to this work. G.R. is
supported by an Australian Postgraduate Award. D.M.P. acknowledges
support from the Australian Research Council.}

\bibliography{ruben_youngs_pinholes}

\end{document}